# Comment on "Role of spatial coherence in Goos-Hänchen and Imbert-Fedorov shifts"

Li-Gang Wang[*]

*Department of Physics, Zhejiang University, Hangzhou, 310027, China*
*Corresponding author: sxwlg@yahoo.com.cn*

It is shown that the spatial Goos-Hänchen shift is greatly affected by spatial coherence. A typical example is given.

A recent paper [1] presented a general formula of both Goos-Hänchen (GH) and Imbert-Fedorov shifts for partially coherent light beams. Their result shows that the spatial GH shift does not depend on the spatial coherence, see Eq. (12) or (20) in Ref. [1]. In the derivations, the authors actually made a first-order Taylor expansion on the reflection coefficient around a angle $\theta_0$, similar to Ref. [2]. This is incorrect for partially coherent beams.

Let us consider a Gauss-Schell beam (GSB) reflected from an interface, as the same as in Ref. [1]. For simplicity, we only consider one transverse coordinate, so the cross-spectral density of the GSB with a incident angle $\theta_0$ has the form [3]:

$$W^{(0)}(y_1, y_2) = [S(y_1)S(y_2)]^{1/2} g(\delta), \quad \text{with} \quad \delta = y_2 - y_1,$$

$$S(y) = A^2 \exp[-\frac{y^2}{2\sigma_y^2}], \quad g(\delta) = \exp[-\frac{\delta^2}{2\sigma_{gy}^2}]\exp[-ik_{y0}\delta].$$

Here $k_{y0} = k\sin\theta_0$, $\sigma_y = \sigma_0/\cos\theta_0$, $\sigma_{gy} = \sigma_g/\cos\theta_0$, and these symbols are the same as in Ref. [3]. For its angular spectrum of $W^{(0)}(y_1, y_2)$, it is computed easily and is given by [2, 3]

$$W^{(0)}(k_{y_1}, k_{y_2}) = 2A^2 \sigma_y \tilde{\sigma} \exp\left\{-\frac{(k_{y_1} - k_{y0})^2 + (k_{y_2} - k_{y0})^2}{(1/\tilde{\sigma}^2)}\right\} \times \exp\left[-\frac{(k_{y_1} - k_{y_2})^2}{\sigma_{gy}^2/(2\tilde{\sigma}^2\sigma_y^2)}\right], \quad (1)$$

where $\frac{1}{\tilde{\sigma}^2} = \frac{1}{\sigma_y^2} + \frac{4}{\sigma_{gy}^2}$, $k_{y1,2} = k\sin\theta_{1,2}$. Assume $r(k_y)$ be the reflection coefficient, so that the reflected field at the interface is given by

$$W^{(r)}(y_1, y_2) = \frac{1}{2\pi}\iint W^{(0)}(k_{y_1}, k_{y_2})r(k_{y_1})r^*(k_{y_1}) \\ \times \exp(ik_{y_1}y_1 - ik_{y_2}y_2)dk_{y_1}dk_{y_2}. \quad (2)$$

For a coherent beam ($\sigma_g$ or $\sigma_{gy} \to \infty$), one can do the first-order Taylor expansion on $r(k_y)$ around $k_{y0}$, i.e., $\theta_0$, as follows

$$r(k_y) = r(k_{y0})\exp[(\Theta - i\Delta)(k_y - k_{y0})], \quad (3)$$

as long as $\sigma_0$ (or $\sigma_y$) $\gg \lambda$, where $\lambda$ is the wavelength, $\Theta = \frac{1}{|r(k_{y0})|}\frac{d|r|}{dk_y}\Big|_{k_y=k_{y0}}$ and $\Delta = -\frac{d\phi}{dk_y}\Big|_{k_y=k_{y0}}$ are, respectively, the angular and spatial GH shifts [1], here $\phi$ is the phase of $r(k_y)$. But, for a partially coherent beam, its effective angular spectral width is proportional to $1/\tilde{\sigma}$. In Ref. [3], we have pointed out that as the spatial coherence ($\sigma_g$) decreases, the angular spectrum become broader, also see Section 5. 6. 4 in Ref. [4]. Therefore the first-order Taylor expansion, Eq. (3), is not valid for the partially coherent beams, especially for the cases of the reflected phase $\phi$ with rapid changes. For obtaining the correct result, one has to do the numerical integral (2) in a practical case. Fig. 1 shows a typical effect of spatial coherence on the GH shift of a TM-polarized GSB reflected from an interface of a semi-infinite medium with the dielectric constant $\varepsilon = 2 + i0.1$. It is clear seen that the lateral position of the intensity peak does strongly depend on the spatial coherence.

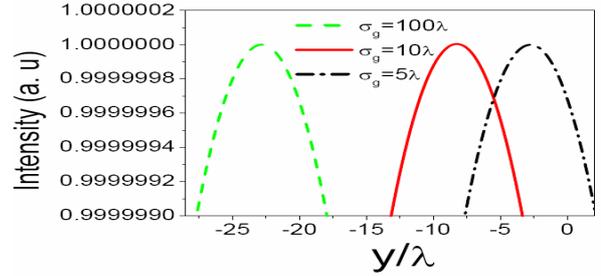

Fig. 1. (Color online). A typical spatial GH shift under different spatial coherence ($\sigma_g$). Other parameters are $\lambda = 500$ nm, $\sigma_0 = 1000\lambda$, and $\theta_0 = 54.7°$ near the Brewster angle.

This work was supported by National Natural Science Foundation of China (No. 61078021) and the National Basic Research Program of China (Grant No. 2012CB921600).